\documentclass[aps,a4paper,superscriptaddress,nofootinbib,onecolumn]{revtex4-1}
\usepackage{amsmath,amssymb,graphicx,bm,multirow}
\usepackage[colorlinks,linkcolor=blue,citecolor=blue,urlcolor=blue ]{hyperref}
\usepackage{acronym}

\usepackage{ulem} 

\newcommand{\be}{\begin{equation}}
\newcommand{\ee}{\end{equation}}
\newcommand{\bea}{\begin{eqnarray}}
\newcommand{\eea}{\end{eqnarray}}
\newcommand{\nn}{\nonumber}

\newcommand{\td}{\tilde}

\newcommand{\TSS}{MOE Key Laboratory of TianQin Mission, TianQin Research Center for Gravitational Physics
$\&$ School of Physics and Astronomy, Frontiers Science Center for TianQin, Gravitational Wave Research Center of CNSA, 
Sun Yat-sen University (Zhuhai Campus), Zhuhai 519082, China}

\newcommand{\PU}{Astronomy Department, School of Physics, Peking University, Beijing 100871, China}
\newcommand{\Kavli }{Kavli Institute for Astronomy and Astrophysics, Peking University, Beijing 100871, China}
\newcommand{\CU}{Institute of Astronomy, University of Cambridge, Madingley Road, Cambridge CB3 0HA, UK}
\def\apj{Astrophysical Journal}  
\def\apjl{Astrophysical Journal} 

\def\apjs{ApJS}                  
\def\mnras{MNRAS}                
\def\prd{Phys.~Rev.~D}           
\def\prl{Phys.~Rev.~Lett}        
\def\aap{A\&A}                   
\def\pasp{PASP}                  

\def\jcap{JCAP}

\allowdisplaybreaks

\begin{document}
\title{An Opacity-Free Method of Testing the Cosmic Distance Duality Relation Using Strongly Lensed Gravitational Wave Signals}

\author{Shun-Jia Huang}
\email{corresponding author: huangshj9@mail2.sysu.edu.cn}
\affiliation{School of Science, Shenzhen Campus of Sun Yat-sen University, Shenzhen 518107, China}
\affiliation{\TSS}

\author{En-Kun Li}
\email{corresponding author: lienk@mail.sysu.edu.cn}
\affiliation{\TSS}

\author{Jian-dong Zhang}
\affiliation{\TSS}

\author{Xian Chen}
\affiliation{\PU}
\affiliation{\Kavli}

\author{Zucheng Gao}
\affiliation{\CU}

\author{Xin-yi Lin}
\affiliation{\TSS}

\author{Yi-Ming Hu}
\email{corresponding author: huyiming@sysu.edu.cn}
\affiliation{\TSS}

\date{\today}

\begin{abstract}
The cosmic distance duality relation (CDDR), expressed as $D_L(z) = (1+z)^2 D_A(z)$, plays an important role in modern cosmology.
In this paper, we propose a new method of testing CDDR using strongly lensed gravitational wave (SLGW) signals.
Under the geometric optics approximation, we calculate the gravitational lens effects of two lens models, the point mass and singular isothermal sphere. 
We use functions of $\eta_1(z) = 1+\eta_0 z$ and $\eta_2(z) = 1+\eta_0 z/(1+z)$ to parameterize the deviation of CDDR.
By reparameterizing the SLGW waveform with CDDR and the distance-redshift relation, we include the deviation parameters $\eta_0$ of CDDR as waveform parameters. 
We evaluate the ability of this method by calculating the parameter estimation of simulated SLGW signals from massive binary black holes.
We apply the Fisher information matrix and Markov Chain Monte Carlo methods to calculate parameter estimation. 
We find that with only one SLGW signal, the measurement precision of $\eta_0$ can reach a considerable level of 0.5-1.3\% for $\eta_1(z)$ and 1.1-2.6\% for $\eta_2(z)$, depending on the lens model and parameters.

\end{abstract}
\keywords{}

\pacs{}
\maketitle
\acrodef{CDDR}{cosmic distance duality relation}
\acrodef{SLGW}{strongly lensed gravitational wave}
\acrodef{GW}{gravitational wave}
\acrodef{EM}{electromagnetic}
\acrodef{SGL}{strong gravitational lensing}
\acrodef{SNe Ia}{Type Ia supernovae}
\acrodef{BAO}{baryon acoustic oscillations}
\acrodef{UCR}{ultra-compact radio}
\acrodef{MBBH}{massive binary black hole}
\acrodef{MBH}{massive black hole}
\acrodef{PM}{point mass}
\acrodef{GOA}{geometric optics approximation}
\acrodef{SIS}{Singular Isothermal Sphere}
\acrodef{QGOA}{quasi-geometric optics approximation}
\acrodef{PN}{post-Newtonian}
\acrodef{FIM}{Fisher information matrix}
\acrodef{MCMC}{Markov Chain Monte Carlo}
\acrodef{SNR}{signal-to-noise ratio}


\tableofcontents

\section{Introduction}\label{sec:1}



The research of modern cosmology heavily relies on the measurement of celestial distances, such as the angular diameter distance $D_A(z)$ and the luminosity distance $D_L(z)$. 
The \ac{CDDR}, as a fundamental relation in modern cosmology, establishes a connection between $D_A(z)$ and $D_L(z)$, that is $D_L(z) = (1+z)^2 D_A(z)$ \cite{Etherington:1933,Ellis:2007}. 
The \ac{CDDR} holds if the spacetime is described by a metric theory of gravity, as long as photons travel along null geodesics, and the photon number is conserved during propagation \cite{Etherington:1933,Ellis:2007}.
Testing the validity of \ac{CDDR} can not only deepen our understanding of the universe but also reveal new physical or astrophysical mechanisms \cite{Bassett:2004,Corasaniti:2006,Ellis:2013}.
Additionally, the \ac{CDDR}, serving as a fundamental relation in cosmology, finds extensive applications in diverse astronomical research. 
These include studies related to the measurements of cosmic curvature through \ac{SGL} systems \cite{Liu:2020,Xia:2017,Qi:2019}, the observations of the large-scale distribution of galaxies and the near-uniformity of the cosmic microwave background radiation temperature \cite{Planck:2020}, as well as investigations into the geometrical shape, gas mass density profile, and temperature profile of galaxy clusters \cite{Cao:2011,Cao:2016,Holanda:2011}.


The idea of testing \ac{CDDR} is simple, just measure $D_A(z)$ and $D_L(z)$ at the same redshift, and then compare these two distances to see whether \ac{CDDR} is valid or not.
Numerous studies have focused on testing \ac{CDDR} using diverse observational datasets (see \cite{Holanda:2010,Li:2011,Liang:2013,Lima:2021,Wu:2015,Ruan:2018,Liao:2022,Li:2018,Xu:2022,Liao:2016,Li:2023,Liu:2023,Lin:2020,Lin:2021,Arjona:2021} and the references therein). 
\ac{SNe Ia} are considered excellent standard candles and are extensively employed for $D_L(z)$ measurements. 
Conversely, $D_A(z)$ is typically derived from a variety of observations such as the Sunyaev-Zeldovich effect of galaxy clusters and the gas mass fraction measurements in galaxy clusters \cite{Holanda:2010,Li:2011,Liang:2013,Lima:2021}, the \ac{BAO} \cite{Wu:2015}, the \ac{SGL} systems \cite{Ruan:2018,Liao:2022}, and the angular size of \ac{UCR} sources \cite{Li:2018}. 
For instance, combining $D_A(z)$ obtained from galaxy clusters with \ac{SNe Ia} provides a means to test \ac{CDDR} \cite{Holanda:2010,Li:2011,Liang:2013,Lima:2021}. 
A new model-independent cosmological test for the \ac{CDDR} was carried out by 
\citet{Xu:2022}, combining the latest five \ac{BAO} measurements and the Pantheon \ac{SNe Ia} sample.
A model-independent methodology employing \ac{SGL} systems and \ac{SNe Ia} to test \ac{CDDR} was proposed by 
\citet{Liao:2016}.
However, the redshift limitation of \ac{SNe Ia} renders \ac{SGL} systems with a source redshift greater than 1.4 impractical for \ac{CDDR} testing, as there are no \ac{SNe Ia} that correspond to these \ac{SGL} systems at the same redshift. 
Consequently, the available data pairs are significantly fewer than the total number of \ac{SGL} systems. 
To make use of the full \ac{SGL} data to test \ac{CDDR}, 
\citet{Li:2023} reconstruct $D_L(z)$ from \ac{SNe Ia} up to the highest redshift of \ac{SGL} using deep learning.
\citet{Liu:2023} present a new method to use the measurements of \ac{UCR} sources and the latest observations of \ac{SNe Ia} to test \ac{CDDR} and use the Artificial Neural Network algorithm to reconstruct the possible evolution of \ac{CDDR} with redshifts.


These traditional methods for testing \ac{CDDR} have some limitations. 
Firstly, distance measurements based on \ac{EM} wave observations are affected by the cosmic opacity \cite{Avgoustidis:2010,Li:2013,Liao:2015}. 
Secondly, it is necessary to assume that the universe is isotropic because these traditional methods measure distances from different objects located at different redshifts and directions, and then use techniques such as interpolation to test \ac{CDDR} \cite{Li:2019}. 
With the implementation of \ac{GW} detection \cite{Abbott:2019,Abbott:2021,Abbott:2021b,Abbott:2021c,Abbott:2021d}, we are now in the era of \ac{GW} astronomy and cosmological research \cite{Abbott:2016,Abbott:2017}. 
Recently, a new method has been discovered that testing \ac{CDDR} using simulated \ac{SLGW} signals based on the third-generation ground-based \ac{GW} detector can avoid these limitations faced by traditional methods \cite{Lin:2020,Lin:2021,Arjona:2021}. 
This is because the propagation of \ac{GW} signals is not affected by the cosmic opacity and can directly measure $D_L(z)$. 
Additionally, if the time delay between the double images of \ac{SLGW} and \ac{EM} observation including the Einstein angle and the velocity dispersion of the \ac{SGL} galaxy are available, $D_A(z)$ to the same object unaffected by the cosmic opacity can be measured.


Future space-based \ac{GW} detectors are also expected to detect \ac{SLGW} signals \cite{Sereno:2010,Gao:2022}.
TianQin is a planned space-based \ac{GW} observatory designed for sensitivity in the millihertz band \cite{Luo:2016}. 
In recent years, considerable efforts have been dedicated to examining and consolidating the scientific potential of TianQin \cite{Hu:2017, Feng:2019, Bao:2019, Shi:2019, Wang:2019, Liu:2019, Fan:2020, Huang:2020, Mei:2021, Zi:2021, Liang:2021, Liu:2022, Zhu:2022a, Zhu:2022b, Zhang:2022, Lu:2022, Sun:2022, Xie:2022, Cheng:2022,Fan:2022, Shi:2023, Ren:2023, Liang:2023, Huang:2023, Lin:2023, Ye:2023, Lyu:2023, Alejandro:2023, Wang:2023, Chen:2023}.
Concerning detected \ac{GW} sources, TianQin's sky localization precision can achieve a range from 1 deg$^2$ to 0.1 deg$^2$ \cite{Wang:2019, Liu:2019, Fan:2020, Huang:2020}, enabling the potential integration of subsequent \ac{EM} observations for multi-messenger astronomy. 
An effective way to obtain the redshift is from the \ac{EM} counterparts.
For example, If the \ac{MBBH} evolve in a gas-rich environment, the gas will be absorbed into the \ac{MBH}, which will generate \ac{EM} radiation across all band \cite{d'Ascoli:2018}. 
As long as the sky localization precision provided by the \ac{GW} signal is good enough, there will be opportunities to realize the detection of \ac{EM} counterparts \cite{Tamanini:2016}.
For a rough estimation of the \ac{SLGW} detection rate, we consider a detection probability of \ac{SLGW} resulting from \ac{MBBH} mergers as approximately 1\% for space-based \ac{GW} detectors \cite{Gao:2022}. Under an optimistic model, the detection rate of \ac{GW} from \ac{MBBH} is estimated to be $\gtrsim O(10^2)/yr$ \cite{Wang:2019}. 
Consequently, the total number of detected \ac{SLGW} events could be as high as $\gtrsim O(5)$ over the five-year mission lifetime.


In this paper, we propose a new method of testing \ac{CDDR} using \ac{SLGW} signals.
By reparameterizing the \ac{SLGW} waveform with \ac{CDDR} and the distance-redshift relation, we include the deviation parameters of \ac{CDDR} as waveform parameters. 
In this way, measurements based on \ac{SLGW} signals can directly constrain the deviation parameters and test \ac{CDDR}.
We assess the ability of this novel method in testing \ac{CDDR} and present preliminary results for TianQin.
We considered \acp{MBBH} as \ac{GW} sources, allowing the potential extension of the redshift range for testing \ac{CDDR} to values significantly greater than the limit imposed by \ac{SNe Ia} observations, which is 1.4.
This new method requires additional \ac{EM} information on the redshifts of the source and lens to test \ac{CDDR}.


This paper is organized as follows. 
In Section \ref{sec:2}, we describe a new method to test \ac{CDDR} using \ac{SLGW} signals.
In Section \ref{sec:3}, we introduce the parameter estimation method used in this study.
In Section \ref{sec:4}, we evaluate the ability to test \ac{CDDR} using \ac{SLGW} signals. 
Finally, we summarize and discuss in Section \ref{sec:5}. 
Throughout this paper, we use $G = c = 1$ and assume a flat $\Lambda$CDM cosmology with the parameters 
$\Omega_M = 0.315$, $\Omega_{\Lambda} = 0.685$ and $H_0 = 67.66 \mathrm{\ km\ s^{-1}\ Mpc^{-1}}$ \cite{Planck:2020}.


%
%

\section{Test CDDR with \ac{SLGW} signals}\label{sec:2}


When the Schwarzschild radius, $R_S\equiv2GM_L/c^2$, of the gravitational lens is much larger than the wavelength of \ac{GW}\footnote{This condition can also be expressed as $M_L\gg10^8 M_{\odot} (f/\mathrm{mHz})^{-1}$.}, the wave effect can be neglected and \ac{GOA} can be applied.
Under the \ac{GOA}, considering the \ac{PM} and \ac{SIS} lens model, the \ac{SLGW} has two images.
The observed \ac{SLGW} signals are given in the stationary phase approximation as \cite{Takahashi:2003,Huang:2023}, 
\bea
{\widetilde{h}}^L\left(f\right)  &=&  \bigg[ |\mu_+|^{1/2} \Lambda(t) e^{ -i(\phi_D + \phi_{p})(t(f)) } \nn\\
&& -i|\mu_-|^{1/2} e^{2\pi i f t_{d}} \Lambda(t(f)+t_{d})  e^{ -i(\phi_D + \phi_{p})(t(f)+t_{d}) } \bigg]  \nn\\
&&   \times \mathcal{A} f^{-7/6} e^{i\Psi(f)} ,
\label{eq:lensed waveform}
\eea
where the GW amplitude and phase are 
\bea
\mathcal{A}= \sqrt{\frac{5}{96}} \frac{\pi^{-2/3}\mathcal{M}_z^{5/6}}{D_L^S}
\label{eq:amp}
\eea
and 
\bea
\Psi(f) = 2 \pi f t_\mathrm{c} - \phi_\mathrm{c} - \frac{\pi}{4} + \psi_{\mathrm{PN}}.
\label{}
\eea
In the above expression, $D_L^S$ is the luminosity distance to the \ac{GW} source, $t_\mathrm{c}$ and $\phi_\mathrm{c}$ are coalescence time and phase, respectively.
$\psi_{\mathrm{PN}}$ is the \ac{PN} phase as a function of the redshifted chirp mass $\mathcal{M}_z=(1+z_S)(M_1 M_2)^{3/5}/(M_1+M_2)^{1/5}$ and symmetric mass ratio $\eta=(M_1 M_2)/(M_1+M_2)^2$, where $M_1$  and $M_2$ are binary masses.
In this work, we expand the \ac{PN} phase and $t(f)$ to 2 PN order (see \cite{Feng:2019} and the references therein).
The detector response, the polarization phase, and the Doppler phase can be defined by
\bea
\Lambda(t)  &=&  \sqrt{ (1+\cos^2\iota )^2{F^+(t)}^2+4\cos^2\iota {F^\times(t)}^2}, \\
\phi_{P}(t)     &=& {\tan}^{-1}\ \left[\frac{{2\cos\iota\ F}^\times(t)}{{(1+\cos^2\iota)F}^+(t)}\right],
\eea
and
\bea
\phi_D (t) = 2\pi fR\sin\left(\frac{\pi}{2}-\beta\right)\ \cos(2\pi f_\mathrm{m}t-\lambda),
\eea
where $R = 1$ AU, $f_m = 1/\mathrm{year}$ and $\iota$ is inclination angle.
$F^{+, \times}(t)$ is the antenna pattern functions of the source's detector coordinates ($\theta_S, \phi_S$)
 and polarization angle $\psi_S$, and related to the motion and pointing of the detector.
The antenna pattern functions are given by 
\bea
F^+(t,\theta_S,\phi_S,\psi_S) &=& \frac{\sqrt{3}}{2}\left[\frac{1}{2}(1+\cos^2\theta_S)\cos2\phi_S\cos2\psi_S -\cos\theta_S\sin2\phi_S\sin2\psi_S\right]\ ,\nn\\
 F^\times(t,\theta_S,\phi_S,\psi_S) &=& \frac{\sqrt{3}}{2}\left[\frac{1}{2}(1+\cos^2\theta_S)\cos2\phi_S\sin2\psi_S+\cos\theta_S\sin2\phi_S\cos2\psi_S\right]\ .
    \label{eq:}
\eea
where the factor of $\sqrt{3}/2$ comes from the response change from a 90-degree laser interferometer to a 60-degree laser interferometer and $\phi_S(t)=\phi_{S0}+\omega t$ with $\omega \approx2\times10^{-5}$ rad/s being the angular frequency of the TianQin satellites.
The transformation from the source's ecliptic coordinates ($\beta, \lambda$) to the source's detector coordinates for TianQin can be found in Appendix \ref{app:Coordinate transformation}.


For the \ac{PM} lens model used for lensing by compact objects such as black holes or stars, the magnification factor of each image and the time delay between the double images in Eq. (\ref{eq:lensed waveform}) are given by \cite{Takahashi:2003}
\bea
\mu_{\pm}=\frac{1}{2} \pm \frac{y^2+2}{2 y \sqrt{y^2+4}}
\label{eq:PM magnification factor}
\eea
and 
\bea
t_d&=&4M_L (1+z_L)\left[ \frac{y \sqrt{y^2+4} }{2}+ \ln \left(\frac{\sqrt{y^2+4}+y}{\sqrt{y^2+4}-y}\right)\right], \nn \\
&&
\eea
where $y= \eta_L/(2 \sqrt{M_L}) \sqrt{D_A^L/(D_A^SD_A^{LS})}$ is the dimensionless source position in the lens plane, $\eta_L$,  $M_L$ and $z_L$ are the \ac{GW} source position in the lens plane, the lens mass and redshift, respectively.
Here, $D_A^L$, $D_A^S$, and $D_A^{LS}$  are the angular diameter distances to the lens, the source, and from the source to the lens, respectively.

For the \ac{SIS} lens model used for more realistic lens objects than the \ac{PM} lens, such as galaxies, stars cluster and dark matter halo, there are
\bea
\mu_{\pm}= \pm 1+\frac{1}{y}
\label{eq:SIS magnification factor}
\eea
and
\bea
t_d&=&8M_L (1+z_L)y,
\eea
where $y= \eta_L/4 \pi v_d^2 D_A^{LS}$ and the lens mass inside the Einstein radius is given by $M_L = 4\pi^2v_d^4 D_A^LD_A^{LS}/D_A^S$ in this case, and $v_d$ is the velocity dispersion of lens galaxy. 
For this lens model, If $y\ge1$, a single image of $\mu_+$ is formed.
For both \ac{PM} and \ac{SIS} lens models, when $y\rightarrow \infty$ ($\mu_+=1$ and $\mu_-=0$), Eq. (\ref{eq:lensed waveform}) reduces to the unlensed waveform.



In order to test any deviation from \ac{CDDR}, we parameterize it with $(1+z)^2 D_A / D_L = \eta(z)$ and Taylor-expand $\eta(z)$ in two ways (see \cite{Yang:2013,Holanda:2016,Liao:2019} and the references therein): 

(a) expand with redshift $z$, $\eta_1(z) = 1+\eta_0 z$; 

(b) expand with the scale factor $a = 1/(1+z)$, $\eta_2(z) = 1+\eta_0 z/(1+z)$, which avoids divergences at high-$z$.

We reparameterize the \ac{SLGW} waveform (Eq. (\ref{eq:lensed waveform})) and embed the \ac{CDDR} in the waveform, which allows us to constrain the deviation of CDDR based on waveform analysis.
Use the distance-redshift relationship \cite{Hogg:1999} 
\bea
D_A= \frac{1}{H_0(1+z)}\int_0^z dz^\prime \frac{1}{E(z^\prime)},
\label{eq:distance-redshift}
\eea
to reparameterize the angular diameter distance $D_A^L$, $D_A^S$, and $D_A^{LS}$, where $E(z)=\sqrt{\Omega_M(1_z)^3+\Omega_K(1+z)^2+\Omega_{\Lambda}}$ depicts the background evolution of the Universe.
In this case, Eq. (\ref{eq:lensed waveform}) can be rewritten as
\bea
{\widetilde{h}}^L\left(f\right)  &=&  \bigg[ |\mu_+|^{1/2} \Lambda(t) e^{ -i(\phi_D + \phi_{p})(t) } \nn\\
&& -i|\mu_-|^{1/2} e^{2\pi i f t_{d}} \Lambda(t+t_{d})  e^{ -i(\phi_D + \phi_{p})(t+t_{d}) } \bigg]  \nn\\
&&   \times \sqrt{\frac{5}{96}} \frac{\pi^{-2/3}\mathcal{M}_z^{5/6}}{D_A^S(z_S)(1+z_S)^2\eta(z_S)}  f^{-7/6} e^{i\Psi(f)} .
\label{eq:re-lensed waveform}
\eea
Assuming a flat $\Lambda$CDM model, we can describe a lensed waveform according to the above formulas with a set of 13 parameters: the lens mass $M_L$ (for \ac{PM} lens model) or velocity dispersion $v_d$ (for \ac{SIS} lens model), lens redshift $z_L$, \ac{GW} source position in the lens plane $\eta_L$, deviation parameter of \ac{CDDR} $\eta_0$, \ac{GW} source redshift $z_S$, redshifted chirp mass $\mathcal{M}_z$, symmetric mass ratio $\eta$, inclination angle $\iota$, coalescence time $t_c$, coalescence phase $\phi_c$, \ac{GW} source ecliptic coordinates $\beta$, $\lambda$, and polarization angle $\psi_S$.

\section{Parameter estimation methods}\label{sec:3}

%

To evaluate the potential of measuring the Hubble constant using the \ac{SLGW} signals, we apply both analytical and numerical methods with \ac{FIM} and \ac{MCMC}, respectively.
Under the framework of Bayesian inference, one would like to constrain parameters $\theta$ with data $d$, or to obtain the posterior distribution $p(\theta|d,H)$ under model $H$.
Under the Bayes' theorem,
\bea
p(\theta|d,H)=\frac{p(\theta|H)p(d|\theta,H)}{p(d|H)} ,
\label{eq:Bayes}
\eea
the posterior is the product of the prior $p(\theta|H)$ and the likelihood $p(d|\theta,H)$, normalized by the evidence $p(d|H)$.
The likelihood of the \ac{GW} signal can be written as 
\bea
p(d|\theta,H) \propto \exp\left[- \frac{1}{2}\bigg(d-h(\theta)\bigg|d-h(\theta)\bigg)\right],
\label{eq:likelihood}
\eea
where the inner product $(\cdot|\cdot)$ defined as \cite{Finn:1992,Cutler:1994}
\bea 
(a|b)=4\Re e\int^\infty_0 \mathrm{d}f \frac{\tilde{a}^*(f)\tilde{b}(f)}{\tilde{S}_n(f)},
\label{eq:inner product} 
\eea
and $\tilde{S}_n(f)$ is the one-sided power spectral density, we adopt the formula from \cite{Huang:2020} for the expression of TianQin
\bea \td{S}_n(f) &=&\frac{1}{L^2}\left[\frac{4S_a}{(2\pi f)^4}\left(1+\frac{10^{-4} Hz}{f}\right) +S_x\right]\nn\\ &&\times\left[ 1+0.6\left(\frac{f}{f_*}\right)^2 \right]\,,
\label{eq:S_n^NSA} 
\eea
where $L = \sqrt{3}\times10^5\mathrm{km}$,  $S_x = 1\times10^{-24}\mathrm{m}^2\mathrm{Hz}^{-1}$, $S_a = 1\times10^{-30}\mathrm{m}^2\mathrm{s}^{-4}\mathrm{Hz}^{-1}$, and $f_*=c/2\pi L$. 
In our case we consider the \ac{SLGW} waveform $h(\theta)$ described by Eq. (\ref{eq:re-lensed waveform}).



The \ac{FIM} $\Gamma_{ij}$ is defined as 
\be \Gamma_{ij} = \left(\frac{\partial h}{\partial \theta_i}\bigg|\frac{\partial h}{\partial \theta_j}\right),
\label{eq:FIM} \ee
where $\theta_{i}$ stands for the $i$-th \ac{GW} parameter, 
In the limit that a signal is associated with a large \ac{SNR} $\rho = \sqrt{(h|h)} \gg 1$, 
one can approximate the variance-covariance matrix with the inverse of the \ac{FIM}  $\Sigma = \Gamma^{-1}$, with $\Sigma_{ii}$ describing the marginalized variances of the $i$-th parameter.

We adopt the emcee \cite{Mackey:2013} implementation of the affine invariant \ac{MCMC} sampler.




\section{Results}\label{sec:4}

\begin{table} 
	\centering
	\caption{The parameter value and prior for \ac{SLGW} signal of cases (a), (b), (c) and (d).
}
	\label{tab:Parameters value}
	\renewcommand\arraystretch{1.3} %
		\begin{tabular}{ l  c  c  c  c  c }
		\hline
		Parameter & Case & Value & Prior \\ 
		\hline \hline
		\textbf{Lens parameter} & \\
		\multirow{2}{*}{Lens mass $M_L$} &  (a) &  $1.53 \times 10^{12}\, $M$_{\odot}$ & \multirow{2}{*}{$U(10^{9},10^{13})$\,M$_{\odot}$} \\
		                                                        &  (c) &  $1.53 \times 10^{10}\, $M$_{\odot}$ & \\
		\multirow{2}{*}{Velocity dispersion $v_d$} &  (b) &  405.5\,km/s & \multirow{2}{*}{$U(40,1200)$\,km/s}\\
		                                                                   &  (d) &  128.2\,km/s & \\                                                        
		Lens redshift $z_L$ & & 0.5  &-\\ 
		GW source position $\eta_L$ & & 5\,kpc  &$U(0,10)$\,kpc\\ 			
		\hline			
		\textbf{\ac{GW} source parameter} &\\
		Deviation parameter $\eta_0$& & 0& $U(-1,1)$\\
		Source redshift $z_S$ & &1  &-\\  
		Symmetric mass ratio $\eta$ & &0.222 & $U(-1,1)$\\ 
		Inclination angle $\iota$ &  & $\pi/3$   &$U(0,1)$\\  
		Coalescence time $t_c$ &  & $2^{18}$\, s($\sim$3 days)   & $U(10^{5},10^{6})$\, s\\  
		Coalescence phase $\phi_c$ & & $\pi/2$  &$U(-\pi,\pi)$\\  			
		Ecliptic  latitude $\beta$ & & $25.3^{\circ}$ & -\\  
		Ecliptic longitude $\lambda$ & & $120.4^{\circ}$  & -\\  
		Polarization angle $\psi_S$ & & $\pi/2$ & $U(0,\pi)$\\  
		\hline
	\end{tabular}
\end{table}

In order to verify the feasibility of the new method for testing \ac{CDDR} using \ac{SLGW} signals, we present the expected measurement precision of the deviation parameter $\eta_0$ for \ac{CDDR}, based on simulated \ac{SLGW} signals and the instrument specifications of the space-based \ac{GW} detector, TianQin. 
The simulated \ac{GW} source in this paper is \ac{MBBH}.
In this paper, we have considered different cases to compare the measurement precision of the deviation parameter $\eta_0$, i.e. (a) \ac{PM} model with $y=0.3$, (b) \ac{SIS} model with $y=0.3$, (c) \ac{PM} model with $y=3.0$, and (d) \ac{SIS} model with $y=3.0$, different lens models with vary lens masses or velocity dispersions. 

\begin{figure*}[h]
    \includegraphics[width=0.45\textwidth]{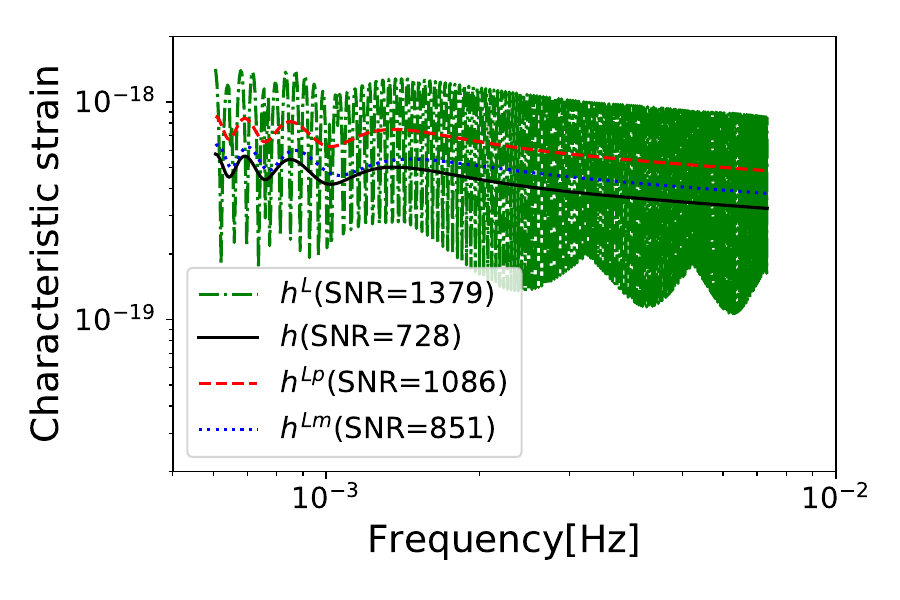}
    \includegraphics[width=0.45\textwidth]{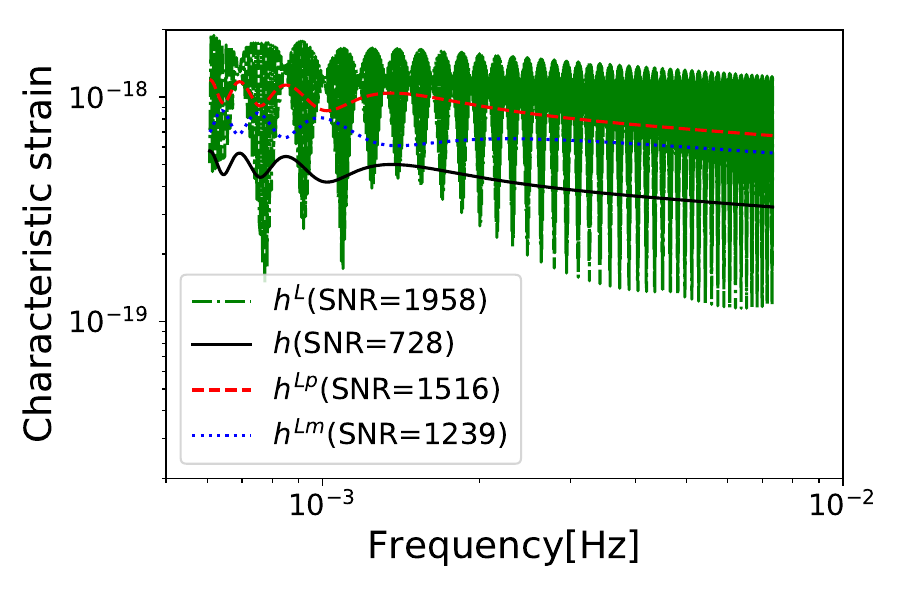}
    \\
    \includegraphics[width=0.45\textwidth]{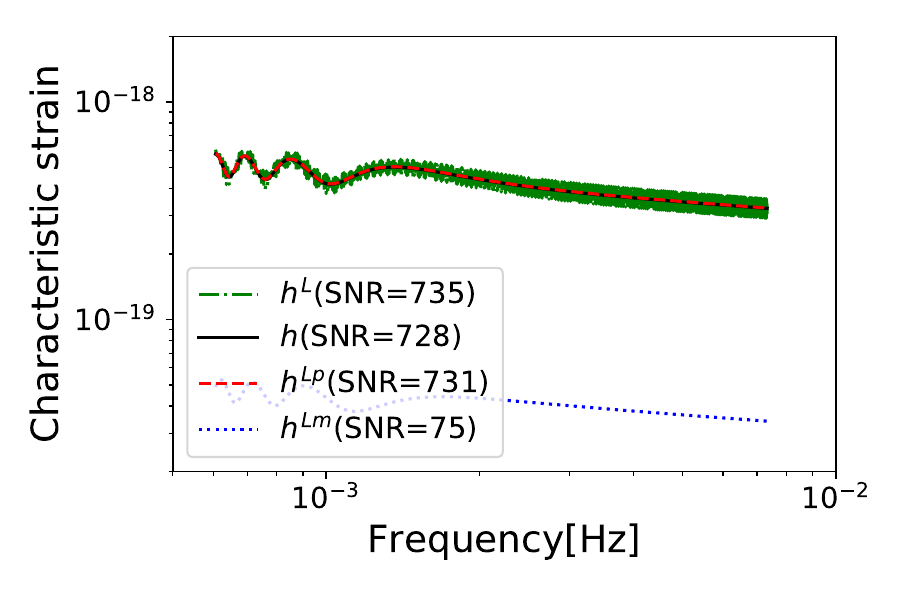}
    \includegraphics[width=0.45\textwidth]{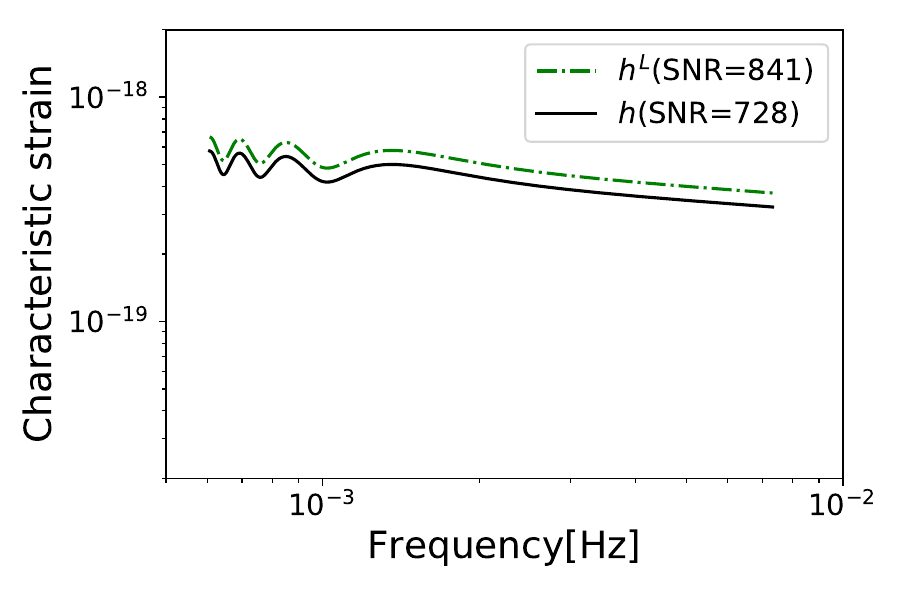}
    \caption{Characteristic strain $h_c=f\widetilde{h}^L(f)$ of \ac{SLGW} signals for four cases:
The upper left corner corresponds to the case (a) \ac{PM} model with $y=0.3$;
The upper right corner corresponds to the case (b) \ac{SIS} model with $y=0.3$;
The lower left corner corresponds to the case (c) \ac{PM} model with $y=3.0$;
The lower right corner corresponds to the case (d) \ac{SIS} model with $y=3.0$.
The black solid line in the figure represents the unlensed \ac{GW} signal, while the red dashed line and blue dotted line correspond to the lensed \ac{GW} signal of $\mu_+$ and $\mu_-$, respectively.
The green dashdot line represents the \ac{SLGW} signal, as described in Eq. (\ref{eq:re-lensed waveform}).
Note that for the case (d) \ac{SIS} model with $y\ge1$, the \ac{SLGW} signal has only one image of $\mu_+$.}
    \label{fig:waveform}
\end{figure*}

Given the parameter value (see Table \ref{tab:Parameters value}), the \ac{SLGW} signals can be simulated according to Eq. (\ref{eq:re-lensed waveform}), as shown in Figure \ref{fig:waveform}. 
The vertical axis in Figure \ref{fig:waveform} represents the dimensionless \ac{GW} characteristic strain $h_c = f \widetilde{h}^L(f)$. 
The upper left corner corresponds to the case (a) \ac{PM} model with $y=0.3$;
The upper right corner corresponds to the case (b) \ac{SIS} model with $y=0.3$;
The lower left corner corresponds to the case (c) \ac{PM} model with $y=3.0$;
The lower right corner corresponds to the case (d) \ac{SIS} model with $y=3.0$.
The black solid line in the figure represents the unlensed \ac{GW} signal, while the red dashed line and blue dotted line correspond to the lensed \ac{GW} signal of $\mu_+$ and $\mu_-$, respectively.
The green dashdot line represents the \ac{SLGW} signal, as described in Eq. (\ref{eq:re-lensed waveform}).
Note that for the case (d) \ac{SIS} model with $y\ge1$, the \ac{SLGW} signal has only one image of $\mu_+$.
Due to $\eta_ 0=0$, the two parameterized forms $\eta_1(z)$ and $\eta_2(z)$ of $\eta(z)$ simulate the same \ac{SLGW} signals.

In order to measure the deviation parameter $\eta_0$ of \ac{CDDR} using \ac{SLGW} signals, it is necessary to assume that the redshift of the source and lens is known, that is, an \ac{EM} counterpart of the \ac{GW} source and lens is required.
By the way, the redshift mentioned in this article is the cosmological redshift and does not take into account the Doppler redshift of the \ac{GW} source.
Meanwhile, in order to simplify the calculation, we also fixed the position parameters $\beta$ and $\lambda$ of the \ac{GW}  source. 
Because it is assumed that there is an \ac{EM} counterpart of the \ac{GW} source, this operation is reasonable.
This paper uses the two methods introduced in Section \ref{sec:3}, the \ac{FIM}, and \ac{MCMC}, to calculate the parameter estimation of \ac{SLGW} signals.

The parameter estimation of \ac{SLGW} signals with four cases (a), (b), (c), and (d) was performed using the \ac{FIM}, and the measurement precision of the deviation parameter $\eta_0$ of the \ac{CDDR} was obtained as listed in Table \ref{tab:sigma_eta0}.
For the case (a), the measurement precision of $\eta_0$ is 0.9\% in the parameterized form $\eta_1(z)$, and 1.9\% in $\eta_2(z)$.
For the case (b), the measurement precision of $\eta_0$ is 0.5\% in the parameterized form $\eta_1(z)$, and 1.1\% in $\eta_2(z)$.
For the case (c), the measurement precision of $\eta_0$ is 1.3\% in the parameterized form $\eta_1(z)$, and 2.6\% in $\eta_2(z)$.
For the case (d), regardless of whether the parameterized form $\eta_1(z)$ or $\eta_2(z)$ is used, the measurement precision of $\eta_0$ cannot be obtained. 
This is because the \ac{SIS} model has only one image with \ac{GOA} when $y\ge1$, and in principle, it cannot recognize the \ac{SLGW} signals in this case, thus unable to measure $\eta_0$.

\begin{table*} 
		\centering
		\caption{The measurement precision of $\eta_0$. 
		}
		\label{tab:sigma_eta0}
		\tabcolsep=1.3cm
		\renewcommand\arraystretch{1.5} %
		\begin{tabular}{ c  c c }
			\toprule
			Case &  $\eta_1(z)$ & $\eta_2(z)$   \\ 
			\hline 
			(a)PM,$y=0.3$         & $\pm$ 0.009   &$\pm$ 0.019   \\ 
			(b)SIS,$y=0.3$         & $\pm$ 0.005   &$\pm$ 0.011   \\ 	
			(c)PM,$y=3.0$         &  $\pm$ 0.013   &$\pm$ 0.026   \\ 			
			(d)SIS,$y=3.0$         &  -  & -  \\ 			
			\hline
		\end{tabular}
\end{table*}

\begin{figure*}
	\begin{minipage}{8.5cm}
    		\vspace{0.1cm}
 		\includegraphics[width=1\textwidth]{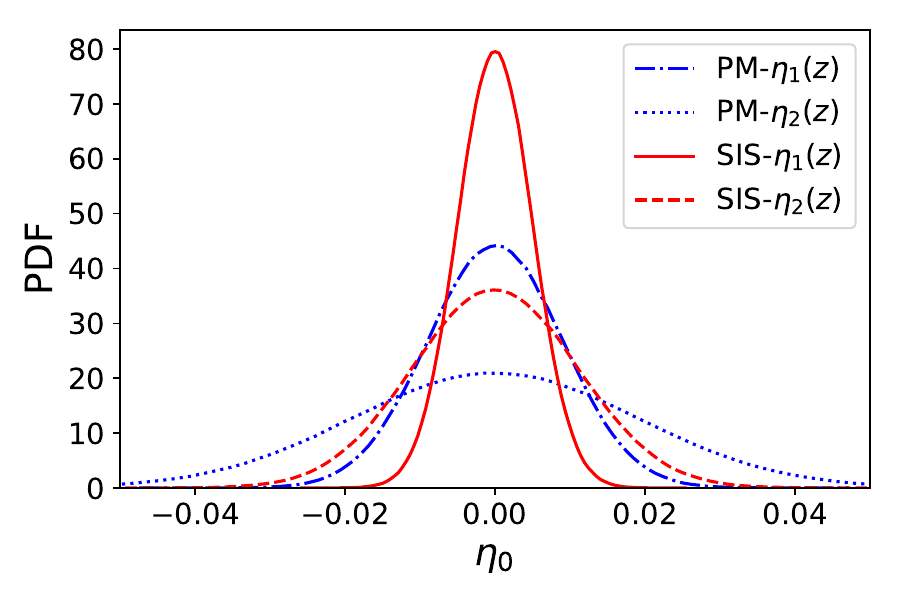}
   	 	\vspace{0.1cm}
  	\end{minipage}
	\begin{minipage}{8.5cm}
    		\vspace{0.1cm}
 		\includegraphics[width=1\textwidth]{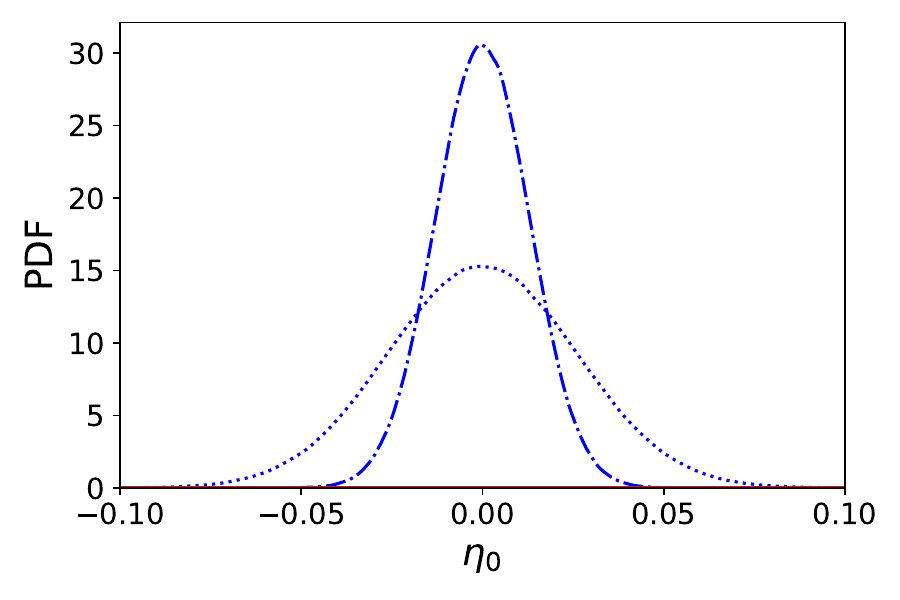}
   	 	\vspace{0.1cm}
  	\end{minipage}
	\caption{
The probability density functions of $\eta_0$ constrained by the \ac{SLGW} signal with two parametric forms: $\eta_1(z) = 1+\eta_0 z$ and $\eta_2(z) = 1+\eta_0 z/(1+z)$.
Left panel: for cases (a) and (b) with $y=0.3$; Right panel: for cases (c) and (d) with $y=3$.
}
	\label{fig:eta0pdf}
\end{figure*}

Assuming that the measurement of the deviation parameter $\eta_0$ follows a normal distribution.
Using $\eta_0=0$ as the mean and the measurement precision of $\eta_0$ listed in Table \ref{tab:sigma_eta0} as the standard deviation, the probability density functions of $\eta_0$ can be drawn, as shown in Figure \ref{fig:eta0pdf}.
The left panel shows the probability density functions of $\eta_0$ for cases (a) and (b) with $y=0.3$.
The blue dashdot and dotted lines correspond to the probability density functions of $\eta_0$ in the parameterized forms $\eta_1(z)$ and $\eta_2(z)$ of the case (a), while the red solid and dashed lines correspond to the probability density functions of $\eta_0$ in the parameterized forms $\eta_1(z)$ and $\eta_2(z)$ of the case (b). 
The right panel shows the probability density functions of $\eta_0$ for cases (c) and (d) with $y=3.0$. 
The blue dashdot and dotted lines correspond to the probability density functions of $\eta_0$ in the parameterized forms $\eta_1(z)$ and $\eta_2(z)$ of the case (c).
In the same probability density functions plot, a distribution that is higher and narrower indicates a higher measurement precision, while a distribution that is shorter and wider indicates a lower measurement precision. 

\boldsymbol{$\eta_1(z)\;\mathrm{vs}\; \eta_2(z)$}\textbf{:}
By comparing the measurement precision of $\eta_0$ in the right two columns of Table \ref{tab:sigma_eta0}, or observing Figure \ref{fig:eta0pdf}, it can be found that for the same case, the parameterized form $\eta_1(z)$ has a better measurement precision of $\eta_0$ than $\eta_2(z)$.
This is determined by the algebraic characteristics of these two parameterized forms, and the parameterized form $\eta_1(z)$ is more sensitive to changes in $\eta_0$ compared to $\eta_2(z)$.
Specifically, the measurement precision of $\eta_0$ for $\eta_1(z)$ is almost half of that for $\eta_2(z)$. 
This is because, in this paper, when the redshift of the \ac{GW} source is $z_S=1$, the deviation term in $\eta_2(z) = 1+\eta_0 z/(1+z)$, $\eta_0 z/(1+z)$, is half of the deviation term in $\eta_1(z) = 1+\eta_0 z$, which is $\eta_0 z$ multiplied by $1/(1+z)=1/2$.
Similarly, if the redshift of the \ac{GW} source is taken as $z_S=2$, then the measurement precision of $\eta_0$ for $\eta_1(z)$ is almost one-third of that for $\eta_2(z)$, which is $1/(1+z)=1/3$ times that of $\eta_2(z)$.

\textbf{\ac{PM} vs \ac{SIS}:}
By comparing the second and third rows of Table \ref{tab:sigma_eta0}, or examining the left plot of Figure \ref{fig:eta0pdf}, it can be found that when $y$ and the parameterized form $\eta(z)$ are the same, the \ac{SIS} model has a better measurement precision of $\eta_0$ than the \ac{PM} model. 
This is because at the same parameters value, the gravitational lensing effect for the \ac{SIS} model is stronger than that for the \ac{PM} model, that is, the magnification factor of the gravitational lensing calculated according to Eq. \ref{eq:SIS magnification factor} is greater than that calculated by Eq. \ref{eq:PM magnification factor}.
Therefore, in the first row of Figure \ref{fig:waveform}, the \ac{SNR} of the \ac{SLGW} for the  \ac{SIS} model is higher at 1958, while the  \ac{SNR} of the \ac{PM} model is 1379.

\boldsymbol{$y=0.3\;\mathrm{vs}\; y=3.0$}\textbf{:}
By comparing the second and fourth rows of Table \ref{tab:sigma_eta0}, or comparing the two subgraphs in Figure \ref{fig:eta0pdf}, it can be found that for the same gravitational lens model, the measurement precision of $\eta_0$ is better for $y=0.3$ compared to when $y=3.0$.
This is because as $y$ decreases, the gravitational lens effect becomes more significant, resulting in an improvement in the measurement precision of $\eta_0$.

\begin{figure*}[h]
    \includegraphics[width=0.45\textwidth]{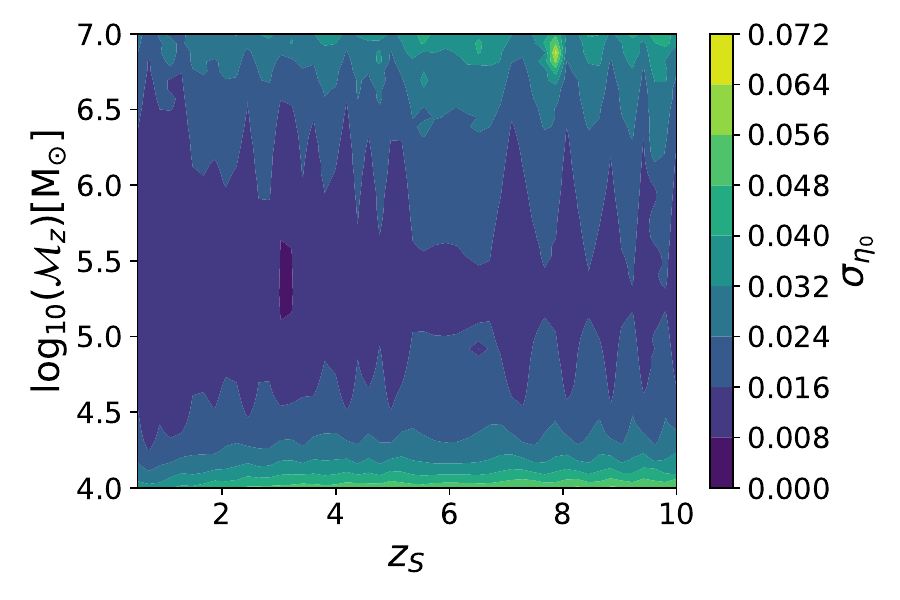}
    \includegraphics[width=0.45\textwidth]{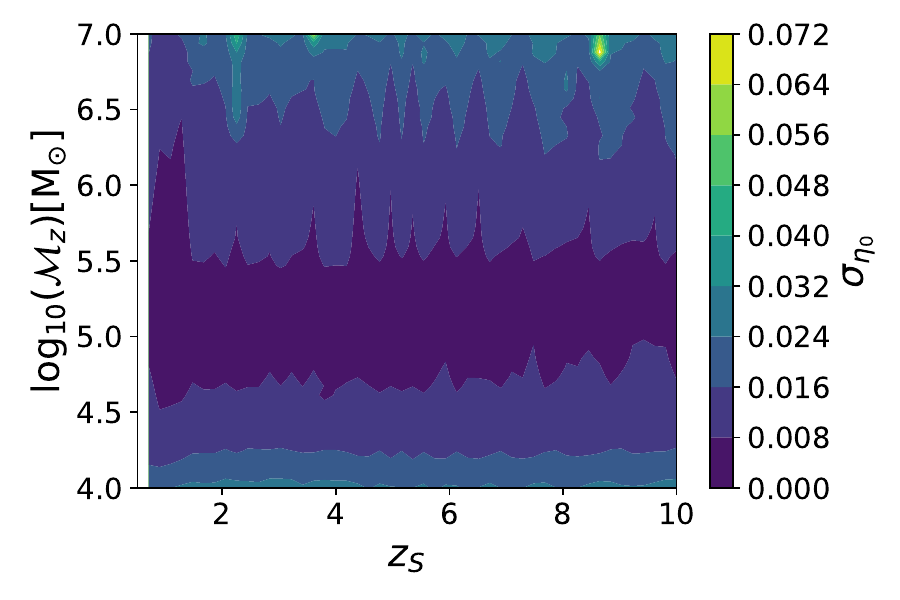}
    \caption{
The contour maps of the measurement precision of $\eta_0$ for both the \ac{PM} model and the \ac{SIS} model.
Left panel: for the \ac{PM} model; right panel: for the \ac{SIS} model. 
The white region in the right panel corresponds to the area where $y\ge1$ in the \ac{SIS} model.
}
    \label{fig:eta0_contour}
\end{figure*}

Using the \ac{FIM} calculation, the contour maps of the measurement precision of $\eta_0$ in the parameterized form $\eta_1(z)$ are shown in Figure \ref{fig:eta0_contour} with the two-dimensional parameter space of the chirp mass and \ac{GW} source redshift for both the \ac{PM} model and the \ac{SIS} model. 
The contour map of the \ac{PM} model is shown on the left panel of Figure \ref{fig:eta0_contour}, while the \ac{SIS} model is shown on the right panel. 
The white region in the right panel corresponds to the area where $y\ge1$ in the \ac{SIS} model.
From Figure \ref{fig:eta0_contour}, it can be seen that both the contour maps of the \ac{PM} model and the \ac{SIS} model exhibit a symmetrical distribution about the chirp mass $\log_{10}(\mathcal{M}_z)=5.3$ M$\odot$. 
At the same redshift, the closer the chirp mass is to $\log_{10}(\mathcal{M}_z)=5.3$ M$\odot$, the higher the measurement precision of $\eta_0$.
As the redshift changes, the contour map shows oscillatory shapes.
The characteristics of these contour maps in Figure \ref{fig:eta0_contour} are related to two factors: the \ac{SNR} of the \ac{SLGW} signal and the amplification factor of the gravitational lensing effect.

\begin{figure}[h]
    \includegraphics[width=0.45\textwidth]{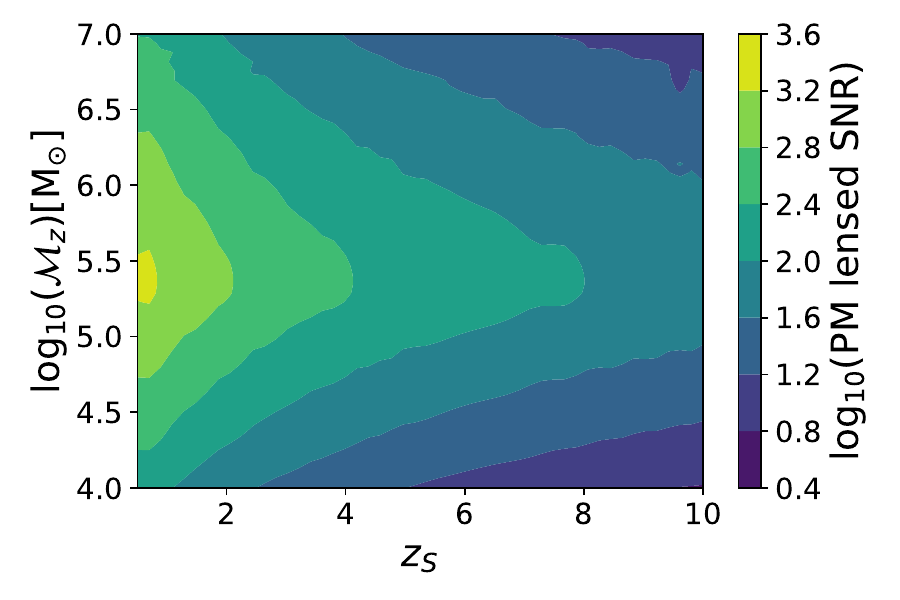}
    \includegraphics[width=0.45\textwidth]{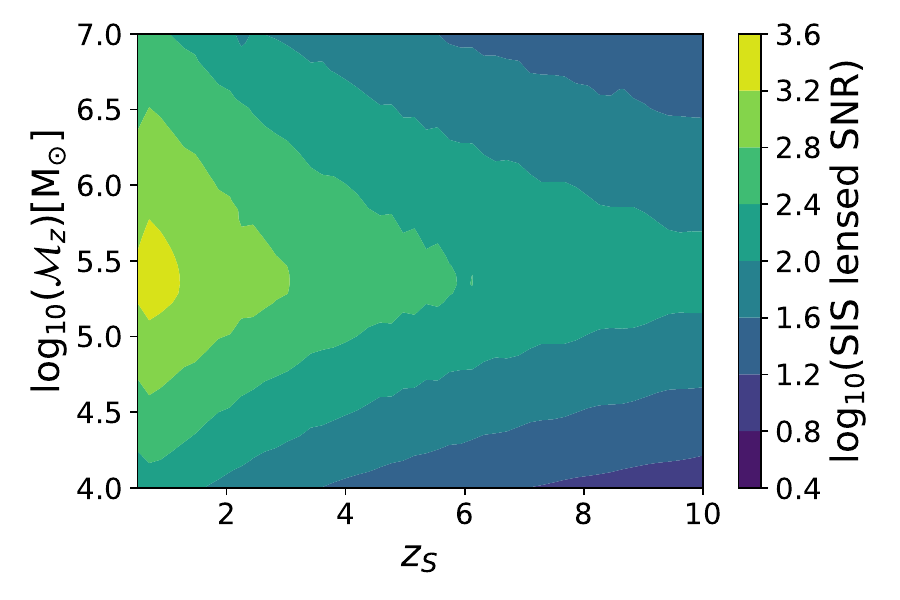} \\
    \includegraphics[width=0.45\textwidth]{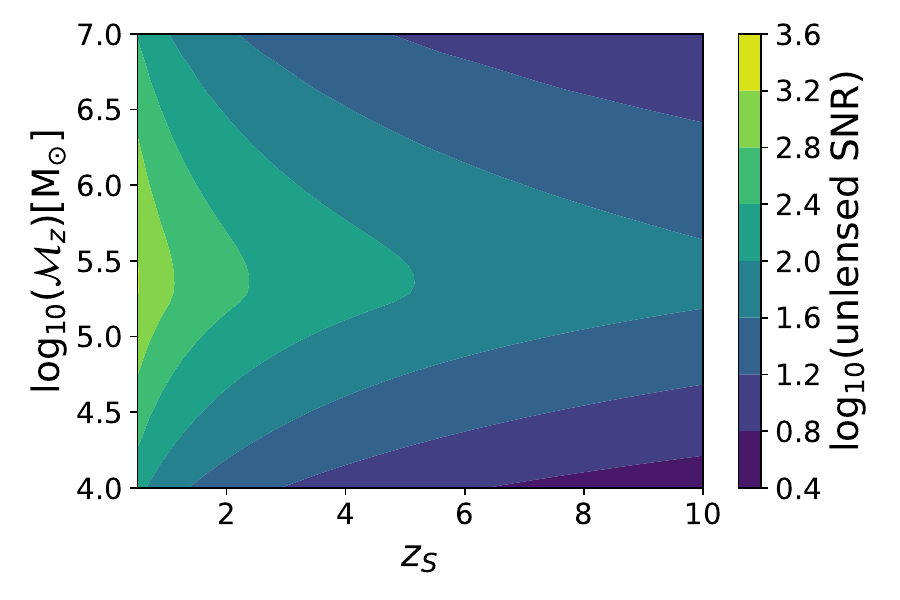}
    \caption{The contour maps of SNR for \ac{GW} signals.
     The upper left panel: the SNR contour map for \ac{SLGW} signals with the \ac{PM} model;
     the upper right panel: the SNR contour map for \ac{SLGW} signals with the \ac{SIS} model;
     the lower panel: the SNR contour map for unlensed \ac{GW} signals. }
    \label{fig:SNR_contour}
\end{figure}

On the one hand, the measurement precision of $\eta_0$ is related to the \ac{SNR} of \ac{SLGW}. 
The contour maps of SNR for \ac{GW} signals are shown in Figure \ref{fig:SNR_contour}, where the upper left panel shows the SNR contour map for \ac{SLGW} signals with the \ac{PM} model, the upper right panel shows the SNR contour map for \ac{SLGW} signals with the \ac{SIS} model, and the lower panel shows the SNR contour map for unlensed \ac{GW} signals. 
All panels show a peak around a small redshift $z_S$ and $\log_{10}(\mathcal{M}_z)=5.3$ M$\odot$. 
Firstly, the smaller the \ac{GW} source redshift $z_S$, the larger the \ac{GW} amplitude as can be seen in Eq. (\ref{eq:re-lensed waveform}), hence the higher the \ac{SNR}. 
Secondly, at the same redshift, the \ac{GW} source with a chirp mass of $\log_{10}(\mathcal{M}_z)=5.3$ M$\odot$ has the highest \ac{SNR} due to the sensitivity band of TianQin. 
This explains the symmetrical distribution in Figure \ref{fig:eta0_contour} concerning the straight line of a chirped mass $\log_{10}(\mathcal{M}z)=5.3$ M$\odot$, and the closer to this line, the better the measurement precision of $\eta_0$.
Therefore, the higher the \ac{SNR} of \ac{SLGW}, the better the measurement precision of $\eta_0$.
When the redshift and chirp masses are the same, the \ac{SNR} of \ac{SLGW} in the \ac{SIS} model is higher than that in the \ac{PM} model. 
As mentioned earlier, this is because the \ac{SIS} model has a stronger gravitational lensing effect.

\begin{figure*}[h]
    \includegraphics[width=0.45\textwidth]{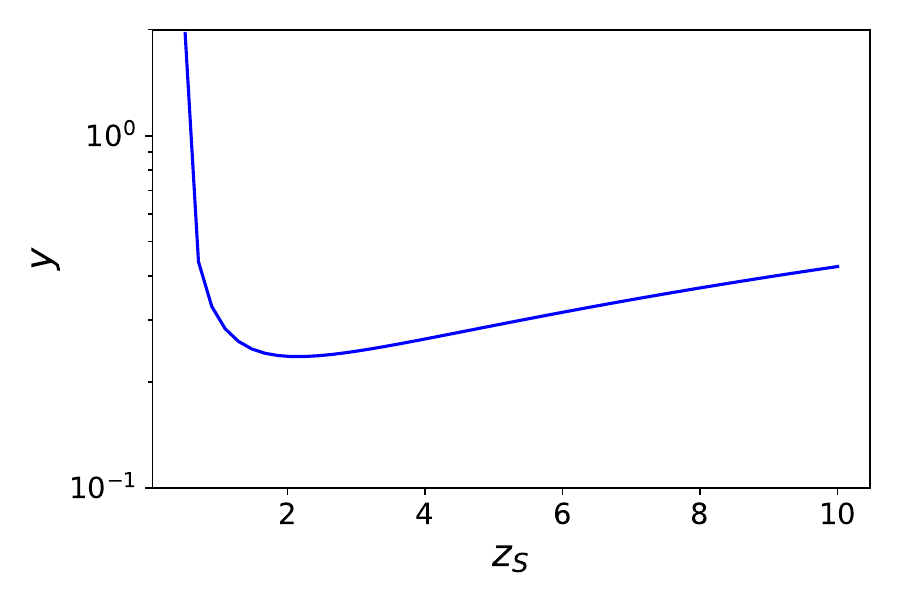}
    \includegraphics[width=0.45\textwidth]{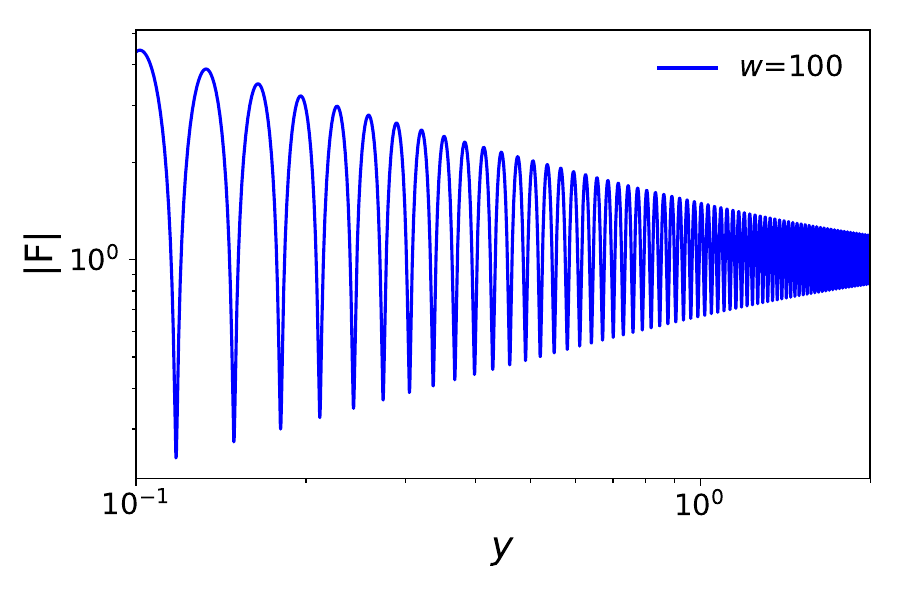}
    \caption{
The graph of $y$ changing with the redshift $z_S$ and the amplification factor $F_{geo}(w,y) = |\mu_+|^{1/2}-i |\mu_-|^{1/2} e^{i2\pi f t_d}$ changing with $y$ in the \ac{PM} model.
}
    \label{fig:y_F}
\end{figure*}

On the other hand, the measurement precision of $\eta_0$ is related to the amplification factor of the gravitational lensing effect. 
Taking the \ac{PM} model as an example, the graph of $y$ changing with the redshift $z_S$ and the amplification factor changing with $y$ is shown in Figure \ref{fig:y_F}. 
As $y$ is independent of the chirp mass $\mathcal{M}_z$, the left panel in Figure \ref{fig:y_F} shows that the value of $y$ changes with the \ac{GW} source redshift $z_S$. 
The right panel in Figure \ref{fig:y_F} shows that the magnitude of the amplification factor oscillates with the change of $y$. 
These can explain the oscillating structure in Figure \ref{fig:eta0_contour}. 

\begin{figure*}
	\includegraphics[width=\textwidth]{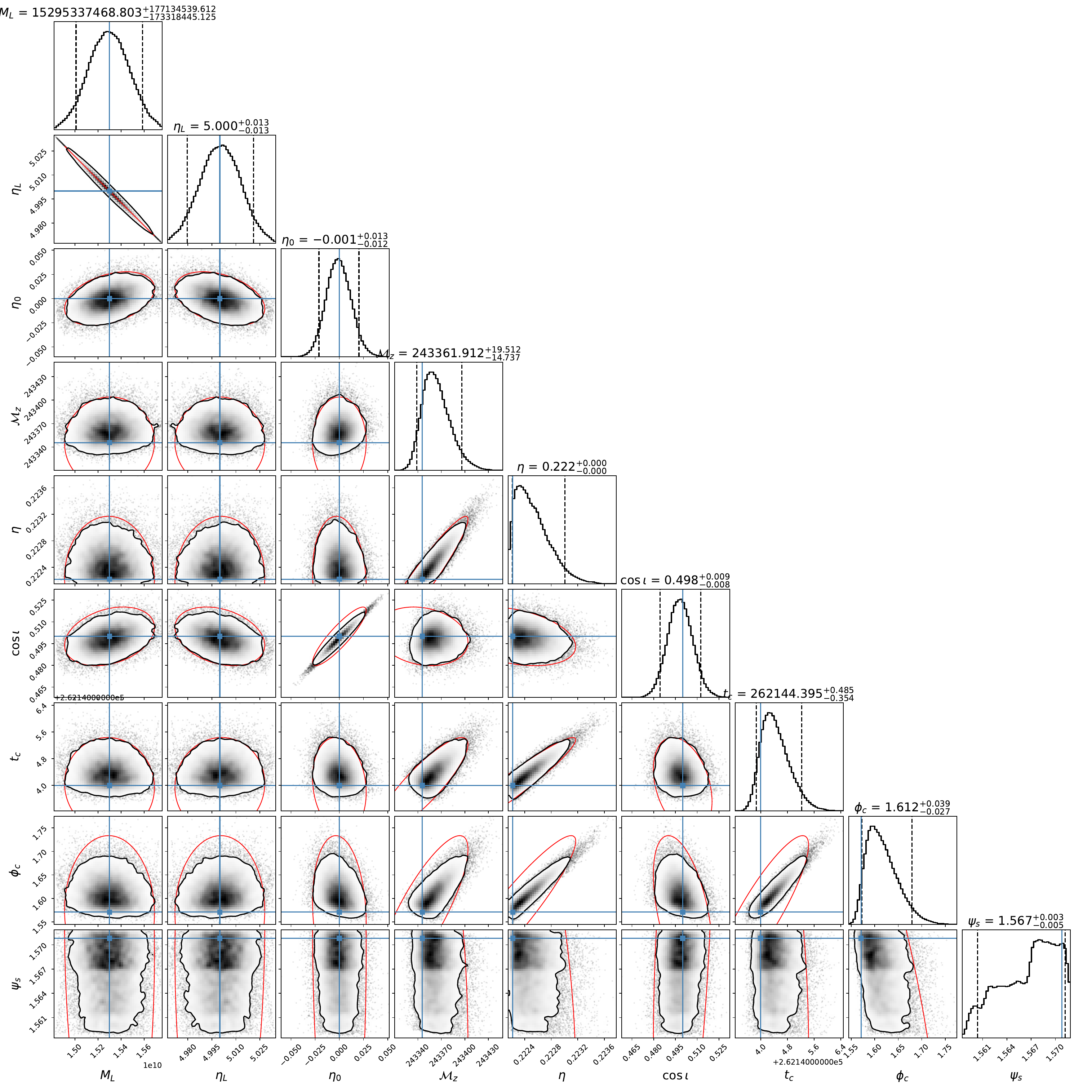}
	\caption{The posterior distributions of the parameters in the parameterized form $\eta_1(z)$ for case (c) \ac{PM} model with $y=3.0$.
   }
	\label{fig:eta1_corner}
\end{figure*}

\begin{figure*}
	\includegraphics[width=\textwidth]{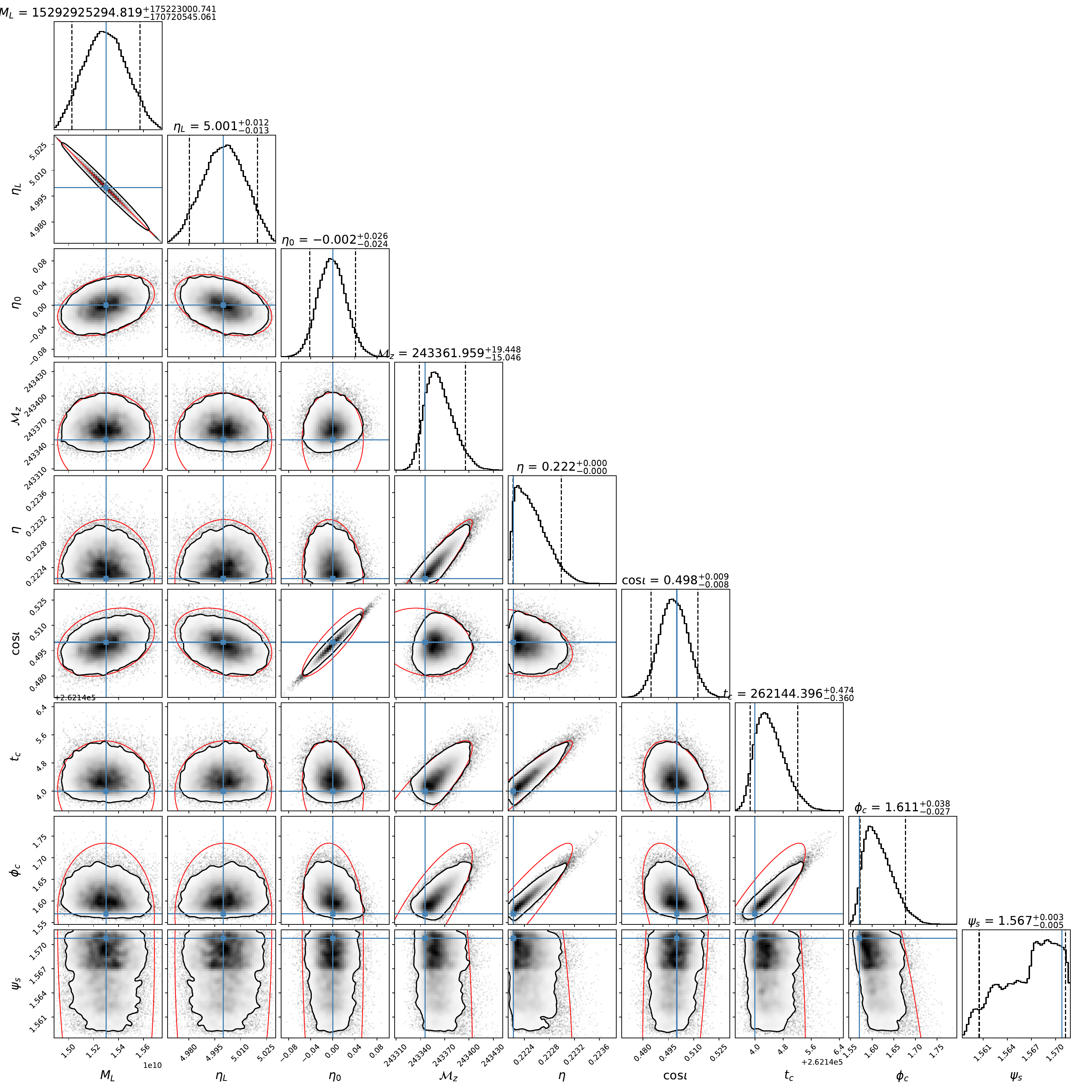}
	\caption{The same as Figure \ref{fig:eta1_corner}, but in parameterized form $\eta_2(z)$.
}
	\label{fig:eta2_corner}
\end{figure*}

When using \ac{MCMC} sampling for parameter estimation, we can set the prior for the parameters shown in Table \ref{tab:Parameters value}. 
For case (c), the posterior distributions of the parameters in the parameterized forms $\eta_1(z)$ and $\eta_2(z)$ are shown in Figures \ref{fig:eta1_corner} and \ref{fig:eta2_corner}, respectively. 
In these figures, the subplots on the diagonal show the histograms of the posterior distributions for each parameter, while the off-diagonal subplots show the two-dimensional density distribution of the 90\% confidence intervals for each pair of parameters with a black line.
The Gelman-Rubin statistic for all parameters' posterior distributions is less than 1.05, which means that the \ac{MCMC} sampling has converged. 
For comparison, we also cover the 90\% confidence intervals calculated by the \ac{FIM} in red ellipses in the same figures. 

It can be found that both the \ac{FIM} and \ac{MCMC} sampling methods give consistent results for parameter estimation. 
The relative measurement errors of each parameter in the parameterized form $\eta_1$ can be obtained by Figure \ref{fig:eta1_corner}: $\sigma_{M_L}/M_L\approx1.2\%$, $\sigma_{\eta_L}/\eta_L\approx0.3\%$, $\sigma_{\mathcal{M}_{z}}/\mathcal{M}_{z}\approx8.0\times10^{-5}$, $\sigma_{\eta}/\eta\approx0.4\%$, $\sigma_{\cos\iota}/\cos\iota\approx1.8\%$, $\sigma_{t_c}/t_c\approx1.8\times10^{-6}$, $\sigma_{\phi_c}/\phi_c\approx1.7\%$, and $\sigma_{\psi_S}/\psi_S\approx0.2\%$; the absolute measurement error of $\eta_0$ is 1.3\%.
From Figure \ref{fig:eta2_corner}, it can be found that in the parameterized form $\eta_2$, the relative measurement errors of each parameter are almost identical to those in $\eta_1$, except for an increase in the absolute measurement error of $\eta_0$ to 2.6\%.





\section{Summary and discussion}\label{sec:5}




This paper aimed to assess the viability of a novel method for testing the \ac{CDDR} using \ac{SLGW} signals. 
The investigation utilized simulated \ac{SLGW} signals and the specifications of TianQin, focusing on a simulated \ac{GW} source involving \acp{MBBH}. 
Four cases, namely (a) \ac{PM} model with \(y=0.3\), (b) \ac{SIS} model with \(y=0.3\), (c) \ac{PM} model with \(y=3.0\), and (d) \ac{SIS} model with \(y=3.0\), were considered to compare the measurement precision of the deviation parameter $\eta_0$ of \ac{CDDR}.

The simulated \ac{SLGW} signals were generated based on the parameters provided in Table \ref{tab:Parameters value}, and their waveforms are depicted in Figure \ref{fig:waveform}. 
Parameterized forms \(\eta_1(z)\) and \(\eta_2(z)\) were employed to simulate \ac{SLGW} signals corresponding to \(\eta_0=0\). 
For testing \ac{CDDR} using \ac{SLGW} signals, the assumption of known source and lens redshifts and fixed position parameters of the \ac{GW} source was made. 
Two methods, \ac{FIM} and \ac{MCMC}, were employed for the parameter estimation of \ac{SLGW} signals.

The \ac{FIM}-based parameter estimation, conducted for the four gravitational lens models, revealed the measurement precision of \(\eta_0\) as listed in Table \ref{tab:sigma_eta0}. 
Case (a) exhibited a precision of 0.9\% for \(\eta_1(z)\) and 1.9\% for \(\eta_2(z)\), while case (b) showed 0.5\% for \(\eta_1(z)\) and 1.1\% for \(\eta_2(z)\). 
For case (c), the precision was 1.3\% for \(\eta_1(z)\) and 2.6\% for \(\eta_2(z)\). 
However, case (d) could not obtain the precision of \(\eta_0\) due to the single image issue.
Assuming a normal distribution for \(\eta_0\) measurements, probability density functions were plotted, demonstrating that \(\eta_1(z)\) generally exhibited better measurement precision than \(\eta_2(z)\). 
Furthermore, comparisons were made for different lens models and \(y\) values, highlighting the impact on precision.
The study delved into the relation between measurement precision of \(\eta_0\) and \ac{SNR} of \ac{SLGW} signals, emphasizing the importance of source redshift and chirp mass. 
It was observed that a higher \ac{SNR} corresponded to better measurement precision. 
Additionally, the effect of the amplification factor in gravitational lensing was discussed.

The consistency of parameter estimation methods between \ac{FIM} and \ac{MCMC} was established. 
For instance, in the case (c), the relative measurement errors for parameters in \(\eta_1(z)\) were approximately \(1.2\%\) for \(M_L\), \(0.3\%\) for \(\eta_L\), \(8.0 \times 10^{-5}\) for \(\mathcal{M}_z\), \(0.4\%\) for \(\eta\), \(1.8\%\) for \(\cos\iota\), \(1.8 \times 10^{-6}\) for \(t_c\), \(1.7\%\) for \(\phi_c\), and \(0.2\%\) for \(\psi_S\), resulting in an absolute measurement error of \(1.3\%\) for \(\eta_0\). 
In \(\eta_2(z)\) case, the absolute measurement error of \(\eta_0\) increased to \(2.6\%\).

Overall, this investigation provided a comprehensive analysis of the proposed method's potential for testing \ac{CDDR} using \ac{SLGW} signals, taking into account various models and parameters. 
The results indicate that the method holds promise for precision measurements on $\eta_0$, particularly when employing certain parameterized forms and gravitational lens models.

In this paper, some assumptions and simplifications were adopted to evaluate the proposed new method of testing \ac{CDDR} using \ac{SLGW} signals.
Firstly, we only calculate the \ac{SLGW} using the gravitational lens models of \ac{PM} and \ac{SIS}, which are applicable to compact objects, stars, galaxies, galaxy clusters, and dark matter halos. 
In the future, we can further extend the gravitational lens models, including more complex models such as the NFW model \cite{Navarro:1996,Navarro:1997}.
Secondly, the \ac{SLGW} waveform used in this paper was given by the \ac{GOA}. 
Although the \ac{GOA} is still valid for the parameter range selected in this paper, the wave optics effect needs to be considered for more general cases \cite{Takahashi:2003}.
Last but not least, in order to save computation time and improve efficiency, we used a \ac{PN} expansion waveform that describes the quasi-circular orbit and no-spin binary  star inspiral phase when calculating the \ac{SLGW} signals from the merger of \ac{MBBH}. 
Although this operation can greatly save computation time, the omission of the merger and ringdown phases means that the information of the \ac{GW} signal is not fully utilized. 
In the future, the IMRPhenom waveform \cite{Khan:2020} that includes the inspiral, merger, and ringdown phases is required to calculate the \ac{SLGW} signals, so that a more comprehensive analysis can be conducted. 
It is expected that using the IMRPhenom waveform can better constrain parameters including the deviation parameter $\eta_0$ of \ac{CDDR} due to more information from the \ac{GW} signal.



\begin{acknowledgments}

This work was supported by the National Natural Science Foundation of China (Grants No. 12173104 and 12261131504), and Guangdong Major Project of Basic and Applied Basic Research (Grant No. 2019B030302001). 
E. K. L. was supported by the Natural Science Foundation of Guangdong Province of China (Grant No. 2022A1515011862).

\end{acknowledgments}

\appendix
\section{Coordinate transformation}\label{app:Coordinate transformation}
The transformation of the source position from the ecliptic coordinates ($\beta,\lambda$) to the detector coordinates ($\theta_S,\phi_{S0}$) for TianQin is described by the following formula:
\begin{equation}      
\left(                 
\begin{array}{c}   
d\sin\theta_S\cos\phi_{S0} \\  
d\sin\theta_S\sin\phi_{S0} \\ 
d\cos\theta_S \\  
\end{array}
\right) =
R_x(\theta = \beta'-90^\circ)R_z(\theta = \lambda'-90^\circ)  
\left(                 
\begin{array}{c}   
d\cos\beta\cos\lambda \\  
d\cos\beta\sin\lambda \\ 
d\sin\beta            \\  
\end{array}
\right),       
\end{equation}
where $d$ is distance, and $\beta'=-4.7^\circ,\lambda'=120^\circ$ are the ecliptic coordinates of TianQin's reference source, J0806.
The rotation matrices are given by
\begin{equation} 
R_x(\theta) =    
\left(                
\begin{array}{ccc}  
1 & 0 & 0\\ 
0 & \cos\theta & \sin\theta\\ 
0 & -\sin\theta & \cos\theta\\ 
\end{array}
\right),
R_z(\theta) =    
\left(                
\begin{array}{ccc}  
\cos\theta & \sin\theta & 0\\ 
-\sin\theta & \cos\theta & 0\\ 
0 & 0 & 1\\ 
\end{array}
\right).                 
\end{equation}

\bibliographystyle{apsrev4-1}
\bibliography{reference}
\end{document}